\documentclass[12pt]{article}
\usepackage{a4}
\usepackage{epsfig}

\newcommand{\jcp} {\ensuremath{J}}

\newcommand{\nub} {\nu_2}

\newcommand{\Uai} {U_{\alpha i}}

\newcommand{\LD} {L_{\Delta}}
\newcommand{\ND} {N_{\Delta}}
\newcommand{\TT} {\ensuremath{\widetilde{T}}}
\newcommand{\PT} {\ensuremath{\widetilde{P}}}
\newcommand{\pt} {\ensuremath{\widetilde{p}}}

\newcommand{\mat}[9] {\left( \matrix{#1 & #2 & #3 \cr
                                     #4 & #5 & #6 \cr 
                                     #7 & #8 & #9 \cr} \right)}
\newcommand{\matTwo}[4] {\left( \matrix{#1 & #2 \cr
                                      #3 & #4 \cr} \right)}

\newcommand{\mt}{$\mu$-$\tau$}

\newcommand{\JJ} {J}
\newcommand{\FF}{{\ensuremath{\cal F}}}
\newcommand{\GG}{{\ensuremath{\cal G}}}
\newcommand{\CC}{{\ensuremath{\cal C}}}
\newcommand{\AC}{{\ensuremath{\cal A}}}
\newcommand{\BB}{{\ensuremath{\cal B}}}
\newcommand{\DD}{{\ensuremath{\cal D}}}
\newcommand{\LL}{{\ensuremath{\cal L}}}
\newcommand{\NN}{{\ensuremath{\cal N}}}

\newcommand{\Sthl}{\ensuremath{S3_{\ell}}}
\newcommand{\Sthn}{\ensuremath{S3_{\nu}}}
\newcommand{\bt}{\ensuremath{\bf 2}}

\newcommand{\Tr}{\ensuremath{{\rm Tr}}}
\newcommand{\Det}{\ensuremath{{\rm Det}}}

\newcommand{\beq}{\begin{equation}}
\newcommand{\eeq}{\end{equation}}
\newcommand{\bea}{\begin{eqnarray}}
\newcommand{\eea}{\end{eqnarray}}

\def\ni{\noindent}
\def\nl{\hfill\break}
\begin{document}
\begin{flushright}
%arXiv:yymm.nnnn [hep-ph]\\
RAL-TR-2007-013\\
October 15th, 2007\\
\end{flushright}
\vspace{2 mm}
\begin{center}
{\Large 
Plaquette Invariants and the Flavour Symmetric\\
Description of Quark and Neutrino Mixings}
\end{center}
\vspace{1mm}
\begin{center}
{P.~F.~Harrison\footnotemark[1]
and D.~R.~J.~Roythorne\footnotemark[2]\\
Department of Physics, University of Warwick,\\
Coventry CV4 7AL. UK}
\end{center}
\begin{center}
{and}
\end{center}
\begin{center}
{W.~G.~Scott\footnotemark[3]\\
Rutherford Appleton Laboratory,\\
Chilton, Didcot, Oxon OX11 0QX. UK}
\end{center}
\vspace{1mm}
\begin{abstract}
\baselineskip 0.6cm
\ni We present a complete set of new flavour-permutation-symmetric mixing observables. We give 
expressions for these ``plaquette invariants'', both in terms of the mixing matrix elements alone, 
and in terms of manifestly Jarlskog-invariant functions of fermion mass matrices. 
While these quantities are unconstrained in the Standard Model, we point out that remarkably, 
in the case of leptonic mixing, the values of most of them are consistent with zero, corresponding 
to certain phenomenological symmetries. We 
give examples of their application to the flavour-symmetric description of both 
lepton and quark mixings, showing for the first time how to construct explicitly 
weak-basis invariant constraints on the mass matrices, for a number of 
phenomenologically valid mixing ansatze.
\end{abstract}
\vspace{-2.0mm}
\begin{center}
{\it To be published in Physics Letters B}
\end{center}
\footnotetext[1]{E-mail:p.f.harrison@warwick.ac.uk}
\footnotetext[2]{E-mail:d.r.j.roythorne@warwick.ac.uk}
\footnotetext[3]{E-mail:w.g.scott@rl.ac.uk}
\baselineskip 0.6cm
\vspace{0.3cm}
\newpage
\ni {\large \bf 1 The Jarlskogian and Plaquette Invariance}
\nl Jarlskog's $CP$-violating invariant, $\jcp$ \cite{JCP:1}, 
is important in the phenomenology of both quarks and leptons. As well as 
parameterising the violation of a specific symmetry, it has two other properties 
which set it apart from most other mixing observables. 
First, its value (up to its sign) is independent of any flavour labels.\footnote{We 
focus on the leptons, although many of our considerations may be 
applied equally well to the quarks. In the leptonic case, neutrino mass eigenstate 
labels $i=1...3$ take the analogous role to the charge $-\frac{1}{3}$ quark flavour 
labels in the quark case. In this sense, we will often use the term ``flavour'' 
to include neutrino mass eigenstate labels, as well as charged lepton flavour labels.} 
Mixing observables are in general dependent on flavour labels, eg.~the 
moduli-squared of mixing matrix elements, $|\Uai|^2$, certainly depend on 
$\alpha$ and $i$. Indeed, $\jcp$ itself is often 
calculated in terms of a subset 
of four mixing matrix elements, namely those forming a given 
plaquette\footnote{We use a cyclic labelling convention such that $\beta=\alpha+1$, 
$\gamma=\beta+1$, $j=i+1$, $k=j+1$, all indices evaluated mod 3.} \cite{BJ}, 
or box \cite{WW}:
\begin{eqnarray}
\jcp={\rm Im}(\Pi_{\gamma k})={\rm Im}(U_{\alpha i}U^*_{\alpha j}U^*_{\beta i}U_{\beta j}).
\label{jarlskogian}
\end{eqnarray}
However, it is well-known \cite{JCP:1} that the value of $\jcp$ 
does not depend on the choice of plaquette (ie.~on its flavour labels, $\gamma$ and $k$ above) - it is 
``plaquette-invariant''. This special feature originates in the fact that \jcp\ is 
{\it flavour-symmetric}, carrying information sampled evenly across the whole 
mixing matrix. We point-out that in fact, {\it any} observable function 
of the mixing matrix elements, flavour-symmetrised (eg.~by summing over both 
rows and columns), and written in terms of the elements of a single plaquette 
(eg.~using unitarity constraints), will be similarly plaquette-invariant. Both its expression
in terms of mixing matrix elements, as well as its value, will be independent of 
the particular choice of plaquette.

The second exceptional property of \jcp\ is that it may be particularly 
simply related to the fermion mass-(or Yukawa) matrices:
\begin{equation}
\jcp=-i\,\frac{{\rm Det}[L,N]}{2\LD \ND}
\label{jarlskogComm}
\end{equation}
where for leptons, $L$ and $N$ are the charged-lepton and neutrino mass 
matrices respectively\footnote{Throughout this paper, $L$ and $N$ are taken to be Hermitian, 
either by appropriate choice of the flavour basis for the right-handed fields, or as the Hermitian squares, 
$MM^{\dag}$, of the relevant mass or Yukawa coupling matrices. The variables $m_{\alpha}$, $m_i$ 
generically refer to their eigenvalues in either case.} 
(in an arbitrary weak basis) and 
$\LD=(m_e-m_{\mu})(m_{\mu}-m_{\tau})(m_{\tau}-m_e)$ (with an analogous 
definition for $\ND$ in terms of neutrino masses and likewise for the 
quarks).

In this paper, we introduce and classify several new plaquette-invariant 
(ie.~flavour-symmetric mixing) observables, which, in common with $\jcp$, 
are independent of flavour labels and may be simply related to the 
mass matrices. Again, in common with $\jcp$, our new observables 
parameterise the violation of certain phenomenological symmetries which 
have already been considered 
significant \cite{SYMMSGENS, DEMOCRACY, MUTAUSYMM} in leptonic 
mixing. In the next section, we define more precisely what we mean by flavour symmetry.

%\newpage
\vspace{0.6cm}
\ni {\large \bf 2 The {\boldmath $\Sthl\times\Sthn$} Flavour Group}
\nl Our starting point is the matrix of moduli-squared of the mixing matrix elements:
\begin{eqnarray}
P
=\mat{|U_{e1}|^2}{|U_{e2}|^2}{|U_{e3}|^2}
     {{|U_{\mu 1}|^2}}{{|U_{\mu 2}|^2}}{{|U_{\mu 3}|^2}}
     {{|U_{\tau 1}|^2}}{{|U_{\tau 2}|^2}}{{|U_{\tau 3}|^2}}.
\label{Pmatrix}
\end{eqnarray}
The $P$-matrix can be simply related to weak-basis invariant functions 
of the fermion mass matrices \cite{HSW06}, 
a feature which we develop later.
Under permutations of the charged lepton flavour labels (ie.~rows), $P$ transforms 
as the 3-dimensional (natural) representation of $S3$ of lepton flavour, \Sthl. 
Similarly, for independent permutations of the neutrino mass eigenstate labels
(ie.~columns, or neutrino ``flavour'' labels, see \hbox{Footnote 1}), $P$ transforms 
as another copy of the natural representation of $S3$, denoted 
\Sthn\ here. Hence, we have a ${\bf 3}\times {\bf 3}$ natural 
representation of the group $\Sthl\times\Sthn$. 
It is well-known that the natural representation of $S3$ is reducible.

We introduce here a convenient parameterisation of the $P$-matrix:
\bea
P=D+\widetilde{P}=
\mat	{\frac{1}{3}}{\frac{1}{3}}{\frac{1}{3}}
    	{\frac{1}{3}}{\frac{1}{3}}{\frac{1}{3}}
	{\frac{1}{3}}{\frac{1}{3}}{\frac{1}{3}}
+\mat	{-w-x}{w}{x}
    		{-y-z}{y}{z}
		{w+x+y+z}{-w-y}{-x-z}.
\label{Ptilde}
\eea
The four parameters $w$, $x$, $y$ and $z$ appearing in the reduced $P$-matrix, $\PT=P-D$ above, 
completely specify the mixing, up to the sign of the $CP$ violation parameter $\JJ$ \cite{HSW06}. 
For example, tribimaximal mixing \cite{TBM:1} corresponds to $w=0$, $x=-1/3$, $y=0$ and 
$z=1/6$, values which are consistent with current neutrino data \cite{FITS:1}. 
We define $\pt$ as the $2\times 2$ plaquette in the top right-hand 
corner\footnote{The top-right (ie.~``$\tau 1$'') plaquette is chosen simply on the grounds that 
its elements correspond to the most directly measured elements of the mixing matrix in both 
the lepton and quark cases.} of \PT:
\vspace{-0.5mm}
\bea
\pt=\matTwo	{w}{x}
    		{y}{z},
\label{wxyz}
\eea
and note that it tranforms under flavour permutations as a ${\bf 2}\times {\bf 2}$ (real) 
irreducible representation of the $\Sthl\times\Sthn$ group.

Our prototype flavour-symmetric mixing observable, $\jcp$, is invariant under 
even permutations of the charged lepton and neutrino flavour labels, 
and flips sign under odd permutations, ie.~it transforms as a 
{$\bf\overline{1}$}$\times${$\bf\overline{1}$} 
representation under $\Sthl\times\Sthn$ (where $\bf\overline{1}$ means the 
``odd'', or alternating, representation of $S3$).
By analogy, we denote as ``flavour-symmetric'', all observables which transform
as (pseudo-)scalars\footnote{We adopt the term (pseudo-)scalars to denote any of the 
{$\bf 1$}$\times${$\bf 1$}, {$\bf\overline{1}$}$\times${$\bf\overline{1}$}, 
{$\bf 1$}$\times${$\bf\overline{1}$} and {$\bf\overline{1}$}$\times${$\bf 1$} representations of
$\Sthl\times\Sthn$.}
under the $\Sthl\times\Sthn$ group. We focus here on 
functions of the mixing matrix elements 
alone (see \hbox{Footnote 8}) which are homogeneous in the $\pt$-matrix,
classifying them as quadratic, cubic etc.~(there are no non-trivial linear plaquette invariants).

The set of polynomials in $w$, $x$, $y$ and $z$ at any given order comprise a 
representation of 
$\Sthl\times\Sthn$ which may be decomposed into irreducible representations.
The set of quadratic polynomials, $(\bt\times\bt)\otimes(\bt\times\bt)$, contains exactly one each of 
{$\bf 1$}$\times${$\bf 1$}, {$\bf\overline{1}$}$\times${$\bf\overline{1}$}, 
{$\bf 1$}$\times${$\bf\overline{1}$} and {$\bf\overline{1}$}$\times${$\bf 1$}, 
as does the cubic $(\bt\times\bt)\otimes(\bt\times\bt)\otimes(\bt\times\bt)$
(for orders $\geq 4$ , there are multiple 1-dimensional representations of each symmetry).
Hence, for orders up to cubic, such (pseudo-)scalar quantities are uniquely 
defined (up to an arbitrary normalisation) by their order in $\pt$ and their symmetry 
under $\Sthl\times\Sthn$. While standard techniques exist \cite{littlewood} for performing 
these reductions, for orders $\leq 3$ the required polynomial forms are anyway easily 
obtained, symmetrising appropriately, eg.~as indicated in \hbox{Section 1}.

\vspace{0.6cm}
\ni {\large \bf 3 New Plaquette Invariant Mixing Observables}
\nl In \hbox{Table \ref{table:values}}, we introduce our new flavour-symmetric 
mixing observables (ie.~plaquette invariants) for order $\leq 3$, 
and summarise the experimental information on each for both leptons and quarks. 
We postpone giving the explicit expressions for them until Section~4.
We normalise the quantities listed in the first column of the table so that their 
maximum value is unity. For comparison, we also include $\jcp$.
\begin{table*}[t]
\begin{center}
\renewcommand{\arraystretch}{1.25} % enlarge line spacing
\small{
\begin{tabular}{|c|c|c|c|c|c|}
\hline
  Observable  &  Order     & Symmetry:            &Theor.& Exptl.~Range & Exptl.~Range  \\
     Name       &   in $\pt$ &  $\Sthl\times\Sthn$  &   Range    & for Leptons   &  for Quarks      \\
\hline
$\FF$ 	& 2 & {$\bf\overline{1}$}$\times${$\bf\overline{1}$} &   $(-1, 1)$   	&  $(-0.14, 0.12)$  	& $(0.893, 0.896)$ \\
$\GG$ 	& 2 & {$\bf 1$}$\times${$\bf 1$}                &   $(0, 1)$   		&  $(0.15, 0.23)$  	& $(0.898, 0.901)$ \\
-- 	& 2 & {$\bf 1$}$\times${$\bf\overline{1}$} &   $0$   		&  -- 	& -- \\
-- 	& 2 & {$\bf\overline{1}$}$\times${$\bf 1$} &   $0$   		&  --  	& -- \\
$\AC$ 	& 3 & {$\bf\overline{1}$}$\times${$\bf\overline{1}$}&   $(-1, 1)$   	&  $(- 0.065, 0.052)$ & $(0.848, 0.852)$ \\
$\CC$ 	& 3 & {$\bf 1$}$\times${$\bf 1$}&   $(-\frac{1}{27}, 1)$  		&  $(-0.005, 0.057)$  & $(0.848, 0.852)$ \\
$6\sqrt{3} \BB$ 	& 3 & {$\bf\overline{1}$}$\times${$\bf 1$} &   $(-1, 1)$   	&  $(-0.49,0.49)$ & $(1.33,1.80)\!\!\times\!\! 10^{-3}$  \\
$6\sqrt{3} \DD$ 	& 3 & {$\bf 1$}$\times${$\bf\overline{1}$} &   $(-1, 1)$   	&  $(0.265,0.675)$ & $(1.11,1.57)\!\!\times\!\! 10^{-3}$  \\
$6\sqrt{3} \jcp$ & inhom. & {$\bf\overline{1}$}$\times${$\bf\overline{1}$} & $(-1, 1)$ &  $(-0.33,0.33)$ & $(2.78,3.48)\!\!\times\!\! 10^{-4}$ \\
\hline
\end{tabular}
\caption{Properties and values of plaquette-invariant observables. The experimentally 
allowed ranges were estimated (90\% CL) from compilations of experimental 
results \cite{FITS:1,CKMUTFIT}, neglecting any correlations between the input quantities.}
\label{table:values}
}
\end{center}
\end{table*}

%\vspace{-2mm}
As is well-known, for quarks and for Dirac neutrinos, four parameters are sufficient 
to completely determine the mixing matrix (also in the Majorana case, as far as flavour
oscillations are concerned). We therefore expect that fixing the values of any 
four independent plaquette invariants must completely determine the 
mixing matrix, up to discrete ambiguities inherent to the built-in flavour symmetry.
A natural set would be $\FF$, $\GG$, $\CC$ and $\AC$, being the lowest-order set possible 
treating leptons and neutrinos (or up- and down-like quarks) symmetrically.
For example, the phenomenologically successful tribimaximal \cite{TBM:1} mixing 
scheme corresponds to the set of constraints 
\hbox{$\FF=\CC=\AC=0$, $\GG=\frac{1}{6}$}, 
constituting the first flavour-symmetric description of exact tribimaximal mixing 
(one could of course substitute the constraint $\jcp=0$ for the condition on $\GG$). 
These constraints determine tribimaximal mixing only up to $(6 \times 6 = 36)$ 
$\Sthl\times\Sthn$ permutations, with the observed mixing breaking the flavour symmetry 
spontaneously \cite{EXTREMISATION}.

We consider also two extremes of mixing. The now-excluded, 
highly symmetric case of trimaximal mixing \cite{TRIMAX}, in which all elements 
of the mixing matrix have magnitude $1/\sqrt{3}$, is given by:
$\FF=\GG=\CC=\AC=0$ (our variables parameterise 
deviations from this unique form). It is somewhat remarkable that the experimental 
values of three of these quantities (\hbox{Table \ref{table:values}}) are 
consistent with zero for the leptons. This near-vanishing of the leptonic observables 
is a flavour-symmetric expression of the presence of large mixing angles in the lepton sector.
By contrast, the case of no mixing corresponds to the constraints
$\FF=\GG=\CC=\AC=1$, and it is notable that for the quarks, 
the experimental values of our variables are all quite close to unity, 
a flavour-symmetric expression of the observed smallness of the quark mixing angles.

\vspace{0.6cm}
\ni {\large \bf 4 Expression and Interpretation of the New Observables}
\nl We give explicit expressions for our new plaquette-invariant observables in terms of 
the mixing parameters $w$, $x$, $y$ and $z$ (as defined in Eq.~(\ref{Ptilde})). 
%appropriate mixing parameters (the $\PT$ matrix or its irreducible equivalent, $\pt$). 
We also give them in terms 
of the fermion mass matrices, to emphasise the analogy with $\jcp$ (cf.~Eq.~(\ref{jarlskogComm})). 
We use a relationship between $\pt$ and the mass matrices, derivable from the relations of 
Jarlskog and Kleppe \cite{JARLSKOGINV}:
\beq
\pt=\widetilde{M_{\ell}}^T\cdot \TT\cdot \widetilde{M_{\nu}}~~~{\rm where}~~~
\TT_{mn}:={\rm Tr}(\widetilde{L^m}\widetilde{N^n}),
\label{Wexpan}
\eeq
and $\widetilde{L^m}:=L^m-\frac{1}{3}\Tr(L^m)$ is the reduced (ie.~traceless) $m$th power of the 
$L$ mass matrix (and similarly for $\widetilde{N^n}$). 
The $2\times 2$ matrix $\TT$ is closely related to the $T$-matrix introduced in \cite{HSW06}, 
and contains complete information about the mixing, assuming the $L, N$-eigenvalues are known.
The transformation 
matrices are given by:\footnote{The columns of  $\widetilde{M_{\ell}}$ and 
$\widetilde{M_{\nu}}$ are labelled by charged lepton labels ($e$ and $\mu$ here)
and neutrino flavour labels (2 and 3 here) respectively, in accordance with our 
choice of plaquette, $\pt$, Eq.~(\ref{wxyz}).}
\bea
\widetilde{M_{\ell}}=\frac{1}{\LD}
\matTwo	{m_{\mu}^2-m_{\tau}^2}{m_{\tau}^2-m_e^2}
    	{m_{\mu}-m_{\tau}}{m_{\tau}-m_e},\quad
\widetilde{M_{\nu}}=\frac{1}{\ND}
\matTwo	{m_3^2-m_1^2}{m_1^2-m_2^2}
    	{m_3-m_1}{m_1-m_2},
\label{PtildeFromTtilde}
\eea
with Det$\widetilde{M_{\ell}}=\LD^{-1}$ etc. We emphasise that although 
we chose $\pt$ (ie.~a particular plaquette of $\PT$) as our $2\times 2$ 
representation of the flavour group, all the following formulae are completely 
independent of this choice.

%\newpage
\vspace{0.4cm}
\ni {\bf 4.1 $\FF$ (Quadratic {$\bf\overline{1}$}$\times${$\bf\overline{1}$})}
\nl Despite the fact that $\FF$ is quadratic in the $P$-matrix, it turns out (perhaps surprisingly) 
to be expressible as the determinant of $P$ (clearly odd under both $\Sthl$ and $\Sthn$):
\begin{equation}
\FF={\rm Det}\,P=3\,(wz-xy)=3\,{\rm Det}\,\pt=3\frac{{\rm Det}\,\TT}{\LD\ND}.
\label{FF}
\end{equation}
We note the striking similarity between the form of $\FF$ 
in terms of mass matrices given by the last equality here, and the 
RHS of Eq.~(\ref{jarlskogComm}).

From Table 1, we see that the data are compatible with $\FF=0$.
We have met this condition before: for non-trivial mass 
spectrum, it is equivalent to the ``determinant condition''
$\ND\FF =0,\label{detCond}$ derived in \cite{EXTREMISATION} 
(which ensures that the mass-constraining Lagrange multipliers can be 
determined in that case). As long as there are no degeneracies in either 
$L$ or $N$, we see from Eq.~(\ref{FF}) that the determinant condition 
may be simply recast as a condition on the mass matrices\footnote{
The determinant condition may equivalently be expressed in terms of 
the original $3\times 3$ $T$-matrix \cite{HSW06} as 
${\rm Det}\,T=0$, or in terms of the determinant 
of a matrix of cubic commutators, see \cite{hawaiitalk}.}: ${\rm Det}\,\TT=0$.

The condition $\FF=0$ is satisfied by 
any mixing matrix having a trimaximally mixed 
row or column, such as, 
eg.~the experimentally viable $\nub=\frac{1}{\sqrt{3}}(1,1,1)^T$, mass 
eigenstate \cite{SYMMSGENS, DEMOCRACY},
%HSNUFACT06}, 
as is readily verified by 
direct substitution, eg.~$w=y=0$ in Eq.~(\ref{FF}). 
The condition is also satisfied by any mixing matrix exhibiting \mt\ reflection 
symmetry ($P_{\mu i}=P_{\tau i},~\forall~i$ \cite{MUTAUSYMM}), 
or indeed by any mixing 
matrix having two rows (resp.~columns) each of whose corresponding 
elements have equal moduli. In particular, the tribimaximal mixing matrix 
\cite{TBM:1} satisfies $\FF=0$, on two counts \cite{SYMMSGENS}, as this mixing 
ansatz has both a trimaximally mixed $\nu_2$ and \mt\ reflection symmetry.

More generally, $\FF$ measures the ``acoplanarity'' of the 
``$P$-vectors'', ie.~when $\FF=0$, the plane defined 
by any pair of $P$-vectors is independent of the choice of pair (in the 
$\nu_1$, $\nu_2$, $\nu_3$ basis, the $e\!-\!\mu$, $\mu\!-\!\tau$ and $\tau\!-\!e$ planes
coincide iff $\FF=0$). Indeed, the $\FF=0$ symmetry is sufficient 
to protect the flavour composition of any remote source against analysis,
since the asymptotic ($L/E$-averaged) matrix of oscillation probabilities,
\hbox{$<\!{\cal P}\!\!>_{\infty}\,=\!P P^T$} \cite{HSW06}, 
cannot be inverted in that case.

%\newpage
\vspace{0.4cm}
\ni {\bf 4.2 $\GG$ (Quadratic {$\bf 1$}$\times${$\bf 1$})}
\nl $\GG$ is quadratic in $\pt$ (and equivalently in $\TT$):
\bea
{\cal G}
%=\frac{1}{2}\sum_{\alpha i}(\PT_{\alpha i})^2
&=&(w + x + y + z)^2 + (w^2 + x^2 + y^2 + z^2) - (wz + xy)\label{GGW}\\
&=&2\,\Tr\, [\pt^T \!\phi\, \pt\phi]
=\frac{2\,{\rm Tr}\,(\TT^T L_G \TT N_G)}{(\LD\ND)^2},
\label{GGT}
\eea
where $\phi\equiv \matTwo{1}{\frac{1}{2}}{\frac{1}{2}}{1}$ is defined by its invariance 
under transformations with all six $2\times 2$ (real) $S3$ 
permutation matrices: $S_i \phi S^T_i=\phi$. $L_G$ is a (symmetric) matrix of 
weak-basis-invariant functions of $L$ depending only on masses, given by:
\begin{eqnarray}
L_G=\frac{3}{2}
\matTwo	{\Tr\,[(\widetilde{L^2})^2]}{-\Tr\,[\widetilde{L}\cdot\widetilde{L^2}]}
		{-\Tr\,[\widetilde{L}\cdot\widetilde{L^2}]}{\Tr\,[\widetilde{L}^2]},
\label{LG}
\end{eqnarray}
where we denote the reduced mass matrix, $\widetilde{L^1}$, as $\widetilde{L}$ 
for simplicity. A similar definition applies for $N_G$ in terms of the $\widetilde{N^n}$. 

Traces (and determinants) of functions of the mass (or Yukawa) matrices 
(eg. $\Tr\,(L^mN^n)$) such as enter the numerators of our expansions, 
eg.~in Eqs.~(\ref{FF}) and (\ref{GGT}) (also Eqs.~(\ref{CCT}) and (\ref{AC}) below), are 
themselves always flavour-invariant ({$\bf 1$}$\times${$\bf 1$} under $\Sthl\times\Sthn$).
In the general case, they depend in a non-trivial way on the mixing matrix elements {\it and} 
the relevant (mass) eigenvalues. By contrast, the particular combinations appearing here
always {\it factorise} into powers of $\LD$ and $\ND$, and the relevant plaquette-invariant 
(which has no dependence on the mass eigenvalues),\footnote{In this sense, we distinguish 
our flavour-symmetric {\it mixing} observables (ie.~plaquette-invariants), 
having no dependence on the mass eigenvalues, 
from the more general class of flavour-symmetric functions of the mass matrices.}
eg.~$(\LD\ND)^2\times \GG$ in Eq.~(\ref{GGT}).
Turning to the denominators (eg.~Eq.~(\ref{GGT})), the discriminant-like factors, 
$\LD^m$ and $\ND^n$
(which cancel the mass-dependence), arise from the structure of the transformation matrices, 
Eq.~(\ref{PtildeFromTtilde}). $\LD$ and $\ND$ transform as {$\bf\overline{1}$} under 
$\Sthl$ and $\Sthn$ respectively, so that the denominator always carries the symmetry 
of the mixing observable.
%(cf.~eg.~$\FF$ and $\GG$, Eqs.~(\ref{FF}) and (\ref{GGT})).

We note that $\GG$ is unique among our set of four $L\!\leftrightarrow\! N$ symmetric plaquette-invariants, 
$\FF$, $\GG$, $\CC$, $\AC$, in having an experimentally allowed range of values for leptons 
(\hbox{Table \ref{table:values}}) which is not consistent with zero, being instead consistent with the 
tribimaximal value, $\GG=\frac{1}{6}\simeq 0.17$. Furthermore, it can  be shown that 
$\GG$ is in fact, the only one of the four which can be non-zero if each of the other three is zero.

The quadratic {$\bf 1$}$\times${$\bf\overline{1}$} and 
{$\bf\overline{1}$}$\times${$\bf 1$} of $\Sthl\times\Sthn$ are both identically zero, so that
in search of additional non-trivial plaquette invariants, we must now move to higher order.

\newpage
\vspace{0.4cm}
\ni {\bf 4.3 $\CC$ (Cubic {$\bf 1$}$\times${$\bf 1$})}
\nl 
$\CC$ is cubic in $\pt$ (and equivalently in $\TT$):
\bea
\CC
%=\frac{3}{2}\sum_{\alpha i}(\PT_{\alpha i})^3
&=&9(xyz + wyz + wxz + wxy) + \frac{9}{2}[xy(x + y) + wz(w + z)]\label{CCW}\\
&=&\frac{3}{2}\frac{\TT_{mn}\,\TT_{pq}\,\TT_{rs}\,\LL_{\CC}^{(mpr)}\,\NN_{\CC}^{(nqs)}}{(\LD\ND)^2}
\label{CCT}
\eea
where the charged lepton mass tensor, $\LL_{\CC}^{(mpr)}$, 
is constructed from flavour-symmetric observables of the lepton mass matrix
($L_m:={\rm Tr}\,L^m$ etc.).
$\LL_{\CC}$ is symmetric in all its indices so that it has only four independent elements:
\bea
\LL_{\CC}^{222}=3,\qquad \LL_{\CC}^{122}=-2L_1,\qquad 
\LL_{\CC}^{112}=\frac{1}{2}(3L_1^2-L_2),\qquad\LL_{\CC}^{111}=L_3-L_1^3,
\label{LLCC}
\eea
with analogous expressions for $\NN_{\CC}$. We note that the pattern 
remarked-on in the previous section indeed continues, the denominator in 
Eq.~(\ref{CCT}) having even powers of both $\LD$ and $\ND$, ensuring 
the overall {$\bf 1$}$\times${$\bf 1$} symmetry under the flavour group.

For lepton data \cite{FITS:1} we see from \hbox{Table \ref{table:values}} 
that $\CC$ is consistent with zero. We note further that any mixing 
scheme having a trimaximally-mixed column (or row) satisfies the condition
$\CC=0$ (in addition to the constraint $\FF=0$ discussed already in Section 4.1). 
Hence, $\CC$ and $\FF$ parameterise deviations from democracy (also called ``magic-square'')
symmetry \cite{DEMOCRACY}, the symmetry which ensures one trimaximally-mixed 
column.

\vspace{0.4cm}
\ni {\bf 4.4 $\AC$ (Cubic {$\bf\overline{1}$}$\times${$\bf\overline{1}$}})
\nl $\AC$ is cubic in $\pt$  (and equivalently in $\TT$):
\bea
\AC
&=&2(w^3 - x^3 - y^3 + z^3) + 3[wx(w - x) + wy(w - y) + yz(z - y)\cr
&+&xz(z - x)+xy(w + z)-wz(x+y)] + \frac{3}{2}[wz(w + z)-xy(x + y)]\\
&=&\frac{81}{2}\frac{\TT_{mn}\,\TT_{pq}\,\TT_{rs}\,\LL_{\AC}^{(mpr)}\,\NN_{\AC}^{(nqs)}}{(\LD\ND)^3}
\label{AC}
\eea
where the charged lepton mass tensor, $\LL_{\AC}^{(mpr)}$, 
is again constructed from flavour-symmetric observables of the lepton mass matrix.
$\LL_{\AC}$ is again also symmetric in all its indices so that it 
is completely determined by four elements as follows:
\bea
\LL_{\AC}^{222}&=&-\Tr\,[\widetilde{L}^3],
\qquad\qquad\,\,\,\,\,\LL_{\AC}^{122}=\Tr\,[\widetilde{L}^2\cdot\widetilde{L^2}],\cr
\LL_{\AC}^{112}&=&-\Tr\,[\,\widetilde{L}\cdot(\widetilde{L^2})^2\,],
\qquad\LL_{\AC}^{111}=\Tr\,[(\widetilde{L^2})^3],
\label{LLAC}
\eea
with analogous expressions for $\NN_{\AC}$ (the $\widetilde{L^m}$ 
were defined just below Eq.~(\ref{Wexpan})). The odd powers of 
$\LD$ and $\ND$ in the denominator of Eq.~(\ref{AC}) ensure the 
required transformation property under the flavour group, as expected.

For lepton data \cite{FITS:1} we see from \hbox{Table \ref{table:values}} 
that $\AC$ is consistent with zero. We note further that any $P$-matrix with two
rows (or two columns) equal (eg.~with \mt-symmetry), satisfies the condition
$\AC=0$  (in addition to the constraint $\FF=0$ discussed already in Section 4.1).
Hence,  $\AC$ and $\FF$ parameterise deviations from 
\mt-symmetry \cite{MUTAUSYMM} and/or any of its permutations. 
A kind of duality is now apparent between this \mt-symmetric mixing scheme 
and the S3 Group mixing (democracy/magic-square symmetry) \cite{DEMOCRACY} scheme, 
with each requiring $F=0$, and additionally $\AC=0$ and $\CC=0$ respectively 
(we encounter a generalisation of this duality later, Eqs.~(\ref{third})-(\ref{equalVert})).

\vspace{0.4cm}
\ni {\bf 4.5 $\BB$, $\DD$  (Cubic {$\bf\overline{1}$}$\times${$\bf 1$}, {$\bf 1$}$\times${$\bf\overline{1}$}) 
and Higher-order Plaquette Invariants}
\nl The set of cubic plaquette invariants is completed by $\BB$ and $\DD$ where:
\bea
\BB & = & 3\sqrt{3}\,[(w^2x + w x^2 - y^2z - z^2y + w x y + w x z - x y z - w y z) \nonumber \\ 
               & & +\,\frac{1}{2}\,(w^2z - w z^2 + y x^2 - y^2x)]\label{BB}\\
&=&-\frac{9\sqrt{3}}{2}\,\frac{\TT_{mn}\,\TT_{pq}\,\TT_{rs}\,\LL_{\AC}^{(mpr)}\,\NN_{\CC}^{(nqs)}}{\LD^3\ND^2}
\label{BBT}
\eea
and $\DD$ is given by exchanging the roles of $L$ and $N$ throughout 
($x\leftrightarrow y$, etc.), and making an overall (conventional) sign flip. 
The tensors $\LL_{\AC}$, $\NN_{\CC}$ etc.~are those already defined above. The denominator, 
$\LD^3\ND^2$ in Eq.~(\ref{BBT}), carries the transformation properties as usual.
Unlike $\FF$, $\GG$, $\CC$ and $\AC$, the observables 
$\BB$ and $\DD$ are not $L\leftrightarrow N$ symmetric 
(or not up $\leftrightarrow$ down symmetric in the quark case). They are also 
not independent of those defined earlier, satisfying the identities:
\bea
\AC^2+\BB^2+\CC^2+\DD^2 & = & \GG(3\FF^2+\GG^2)/2\\
\AC\CC+\BB\DD&=&\FF(\FF^2+3\GG^2)/4.
\label{BDident}
\eea

Plaquette invariants of order $\geq 4$ are not uniquely defined by their 
symmetry under the flavour group. All are however expressible
in terms of those we have already encountered. They may be flavour 
even-even (eg.~$\FF^2$, $\GG^2$, $\GG\CC$ etc.), 
odd-odd (eg.~$\FF\GG$, $\FF\CC$, $\AC\CC$, $\BB\DD$ etc.) 
or odd-even (eg.~$\BB\GG$, $\BB\CC$, $\FF\DD$) etc.
Of particular interest is the square of the $CP$ violation parameter, 
$\jcp^2$, which is even-even, but is not homogeneous in $\pt$.
It may be written in terms of the invariants already discussed:
\beq
18\JJ^2=1/6 - \GG + (4/3)\,\CC - (1/2)\,\FF^2.
\label{jgfc}
\eeq
Its physical range, $0<18\JJ^2<\frac{1}{6}$, implies non-trivial boundaries 
for the space of $\GG$, $\CC$ and $\FF^2$. Clearly, $\JJ^2$ can be expressed in terms 
of mass matrices, either via Eq.~(\ref{jgfc}), or by squaring 
Eq.~(\ref{jarlskogComm}). While plaquette-invariants may be constructed at any order, 
and may be homogeneous or not, we consider those introduced here to be elemental.

\vspace{0.6cm}
\ni {\large \bf 5 Application to Flavour Symmetric Descriptions of Mixing}
\nl Our plaquette invariants may be used to describe fermion mixing in terms which are 
independent of flavour labels, and we have given in Sections 3 and 4, some examples 
for particular lepton mixing schemes. In this section, we expand the list of mixing 
schemes considered, see \hbox {Table~\ref{table:ansatze}}, where we summarise 
the correspondence between these schemes, constraints on our flavour-symmetric 
mixing observables, and the phenomenological symmetries to which they correspond.
\begin{table*}[htb]
\begin{center}
\renewcommand{\arraystretch}{1.25} % enlarge line spacing
\small{
\begin{tabular}{|l|c|c|c|c|l|c|c|c|}
\hline
         Mixing Ansatz           & $\FF$ & $\GG$ & $\CC$ & $\AC$ & Corresponding  & $18\JJ^2$ & \BB & \DD  \\
                                         &           &            &            &            & Symmetries       &      	    &       &          \\
\hline
No Mixing               		&     1   &      1     &     1    & 1 &  \hbox{$\qquad\,\,\,$ --} & 0 &    0    &    0    \\
Tribimaximal Mixing$^*$ \cite{TBM:1} &    0    & $\frac{1}{6}$ & 0 & 0 &  Dem., \mt, $CP$     &    0  &    0    &   $\frac{1}{12\sqrt{3}}$      \\
Trimaximal Mixing  \cite{TRIMAX} &    0    &      0     &     0    & 0 &  Dem., \mt\  & $\frac{1}{6}$ &   0    &    0   \\
S3 Group Mixing$^*$ \cite{SYMMSGENS, DEMOCRACY}    &    0    &     --     &     0    &      --      & Democracy	& -- &   0    &   --    \\
Two Equal $P$-Rows$^*$  \cite{MUTAUSYMM} &    0  &     --     &     --    &       0     & e.g. \mt\  & -- &   0    &   --     \\
Two Equal $P$-Columns 	&    0  &     --     &     --    &       0     & e.g. 1-2  & -- &   --   &   0      \\
Altarelli-Feruglio$^*$	 \cite{AF} &    0  &     --     &     $\frac{6\GG-1}{8}$  &  0  & \mt, $CP$  & 0 &   0    &   --    \\
Tri-$\chi$maximal Mixing$^*$ \cite{SYMMSGENS}   &    0    &      --    &      0    &       0     & Dem.,  \mt\ & -- &   0   &   --   \\
Tri-$\phi$maximal Mixing$^*$ \cite{SYMMSGENS}   &    0    & $\frac{1}{6}$ &     0    & -- & Dem., $CP$& 0 &   0   &   --   \\
Bi-maximal Mixing \cite{BIMAX} &    0    & $\frac{1}{8}$ & $-\frac{1}{32}$ & 0 & $CP$, \mt, 1-2 & 0 &   0    &    0  \\
\hline
\end{tabular}
\caption{Particular mixing schemes and their corresponding descriptions in terms of constraints on 
plaquette invariants, and symmetries. Those marked with an asterisk $(^*)$ are currently 
phenomenologically viable. 
Although the four $L\leftrightarrow N$ symmetric variables, $\FF$, $\GG$, $\CC$ and $\AC$, 
are sufficient, we include $\BB$ and $\DD$ (and $\jcp^2$) for completeness.}
\label{table:ansatze}
}
\end{center}
\end{table*}

%\vspace{-15mm}
Setting the values of $\FF$, $\CC$, $\AC$ 
and $\GG$ or $\JJ^2$ equal to those given in Table~\ref{table:ansatze}, gives for 
the first time, flavour-symmetric statements of the respective mixing schemes.
For example, the constraints $\FF=\CC=\AC=0$ correspond to the \mt-symmetric 
and democratic ``tri-$\chi$maximal'' mixing ansatz \cite{SYMMSGENS}. 
Such constraints are of course readily cast in manifestly weak-basis invariant form 
\cite{weakBasisInv}, using our expressions for our mixing observables in terms of mass matrices.

As well as the conditions summarised in \hbox {Table~\ref{table:ansatze}}, one can 
construct single flavour-symmetric 
constraints corresponding to less restrictive conditions on the mixing matrix, 
which are nevertheless interesting, and phenomenologically viable, eg.:
\bea
8\CC^3-27\FF^2(\CC\GG-\AC\FF)=0 &\Rightarrow& |U_{\alpha i}|^2=\frac{1}{3}~({\rm for~any~particular~}\alpha, i)\label{third}\\
8\BB^3 - 27\FF^2(\BB\GG -\DD\FF)=0 &\Rightarrow& |U_{\alpha i}|^2=|U_{\beta i}|^2 ~({\rm for~any~particular~}\alpha\neq\beta, i)\label{equalVert}.
%8\DD^3 - 27\FF^2(\DD\GG -\BB\FF)=0 &\Rightarrow& |U_{\alpha i}|^2=|U_{\alpha j}|^2 ~({\rm for~any~}\alpha, i\neq j)\label{equalHor},
\eea
The two constraints, $\FF=0$, $\CC=0$, highlighted earlier, corresponding to 
democracy symmetry, are clearly a special case of Eq.~(\ref{third}).
Each of the constraints, Eqs.~(\ref{third})-(\ref{equalVert}), is a ninth-order equation in 
$\pt$, having one solution for each of the nine possible
locations of the constrained element of $U$.

%\newpage
Finally, we give the (two) simultaneous flavour-symmetric constraints which correspond
to ``any element of $U$ equal to zero'' (eg.~$P_{e3}\equiv|U_{e3}|^2=0$, consistent
with the CHOOZ bound \cite{CHOOZ}):
\beq
{\rm Det}K=0;\quad\jcp=0,
\label{zeroelement}
\eeq
where $54\,\Det\,K=2\AC + \FF(\FF^2 - 2\CC - 1)$ and $K$ is the matrix 
of real parts of plaquette-products of $U$ \cite{Kmatrix, HSW06}, ie.~the $CP$-conserving analogue 
of $\jcp$.\footnote{The $\Det\,K=0$ condition, Eq.~(\ref{zeroelement}), may also be written 
in terms of mass matrices using the $Q$-matrix ($Q$ is the matrix of quadratic mass matrix 
commutators related to $K$ by a simple moment transform \cite{EXTREMISATION, HSW06}). 
The condition becomes simply $\Det\,Q =0$.} We point-out that the first of the two conditions 
in Eq.~(\ref{zeroelement}), is in fact consistent with mixing data for {\it both} 
leptons and quarks, this being manifestly so for leptons 
(see eg.~\hbox{Table \ref{table:values}}). In the quark case, for the known values 
\cite{CKMUTFIT} of the 
Wolfenstein parameters, \hbox{$\lambda$, $A$ and $\overline\rho^2+\overline\eta^2$}, $\Det\,K=0$ 
corresponds to $\alpha=(88\pm 1)^{\circ}$, consistent with the latest fits \cite{CKMUTFIT} 
which give $\alpha=(90^{+7}_{-3})^{\circ}$. Indeed, the two constraints 
``$\Det\,K=0$, $\jcp$ small'', together provide a unified and flavour-symmetric, partial 
description of both lepton and quark mixing matrices, being associated with
the existence of at least one small element in each mixing matrix, $U_{e3}$ and $V_{ub}$ 
respectively.

\vspace{0.6cm}
\ni {\large \bf 6 Summary}
\nl We have introduced several mixing observables which, like $\jcp$, 
are (pseudo-)scalars under the flavour-symmetry group $\Sthl\times\Sthn$ (or the analogous
group for the quarks, $S3_U\times S3_D$).
We have shown how, like $\jcp$, they can also be expressed quite simply in terms of 
weak-basis invariant functions of the mass matrices divided by powers of the 
mass matrix discriminants. Our new observables ``measure'' the violations of certain symmetries 
(again in analogy to $\jcp$) associated with the phenomenologically 
successful tribimaximal \cite{TBM:1} scheme. 
It is remarkable that in the case of the leptons, most of these observables are consistent 
with zero, corresponding to the previously identified democracy and \mt\ symmetries.
Such plaquette-invariant observables may be applied to construct explicitly 
flavour-symmetric constraints on the quark and lepton mixing matrices, 
even though the observed mixing matrices are not themselves flavour-symmetric, 
the flavour-symmetry being spontaneously broken. The main result of this paper is the 
set of such constraints, summarised in Table 2, and the corresponding weak-basis invariant 
constraints in terms of the fermion mass matrices.

%\newpage
\nl{\large \bf Acknowledgments}\\
\ni This work was supported by the UK Science and Technology Facilities Council (STFC). 
PFH acknowledges the hospitality of the Centre for 
Fundamental Physics (CfFP) at the Rutherford Appleton Laboratory, and the Particle Physics
Group at the University of Hawaii, Manoa, where some of this work was carried out.
We acknowledge useful comments made by Probir Roy and Sandip Pakvasa and are grateful to
R. Krishnan for independently cross-checking many of the formulae presented herein.

%\newpage

%
\end{document}